\documentclass[secnumarabic,amssymb, nobibnotes, aps, prd, nofootinbib]{revtex4-2}
\usepackage{amsmath} \usepackage{amssymb}
\usepackage{graphicx} 
\usepackage{epstopdf}
\usepackage{xcolor}
\usepackage{amsmath}
\usepackage{amsmath,amssymb,amsthm,amsfonts,mathrsfs,bm,verbatim}
\usepackage{graphicx,subfigure}
\usepackage{indentfirst}
\usepackage[backref]{hyperref}
\usepackage{slashed}
\usepackage{float}

\newcommand{\bea}{\begin{eqnarray}}
\newcommand{\eea}{\end{eqnarray}}

\begin{document}

\title{Quarkonium spin alignment in sheared quark-gluon plasma}
\author{Peiwei Yu}
\email{yupw5@mail2.sysu.edu.cn}
\affiliation{School of Physics and Astronomy, Sun Yat-Sen University, Zhuhai 519082, China} 
\affiliation{Guangdong Provincial Key Laboratory of Quantum Metrology and Sensing, Sun Yat-Sen University, Zhuhai 519082, China}

\author{Shu Lin}
\email{linshu8@mail.sysu.edu.cn}
\affiliation{School of Physics and Astronomy, Sun Yat-Sen University, Zhuhai 519082, China} 
\affiliation{Guangdong Provincial Key Laboratory of Quantum Metrology and Sensing, Sun Yat-Sen University, Zhuhai 519082, China}

\begin{abstract}
We study shear induced correction to in-medium quarkonium potential. The exchanged gluons gain anisotropic Debye mass in quark-gluon plasma with shear, giving rise to a tensor type correction to the quarkonium potential. The correction is calculated in the hard thermal loop approximation, which leads to a D-wave correction to the quarkonium wave function. Application to spin alignment of $J/\psi$ produced in heavy ion collisions is discussed.
\end{abstract}

\maketitle

\section{Introduction}

The past decade has witnessed a growing interest in understanding spin dynamics in relativistic heavy ion collisions (HIC). Since the early proposal of spin polarization for baryon based on spin-orbit coupling \cite{Liang:2004ph,Liang:2004xn}, the first experimental evidence was found in (momentum averaged) global polarization of $\Lambda$ hyepron \cite{STAR:2017ckg}. It has been well described by spin polarization in a vortical quark-gluon plasma (QGP) \cite{Becattini:2013fla,Fang:2016vpj,Huang:2020xyr}. More measurement of global polarization of other baryons have been made \cite{STAR:2020xbm}, which fits qualitatively within the same picture. Momentum dependent local polarization has been measured across different systems \cite{STAR:2019erd,ALICE:2021pzu,CMS:2025nqr}, for which vorticity alone cannot explain even the sign of the measured data \cite{Becattini:2017gcx,Wei:2018zfb,Wu:2019eyi,Fu:2020oxj}. It has been understood now shear of QGP consists an important part in the contribution to local polarization \cite{Fu:2021pok,Becattini:2021suc,Yi:2021ryh}, though uncertainties from collisional effect and baryon structure still remain \cite{Lin:2022tma,Lin:2024zik,Wang:2024lis,Fang:2024vds,Fang:2022ttm}.

The situation is far less clear for spin alignment of vector mesons. The spin alignment measurement in HIC distinguishes from counterpart in proton collisions in the choice of quantization axis: a quantization axis perpendicular to the event plane correlates the mechanism of spin alignment with the unique geometry of HIC \cite{STAR:2022fan,ALICE:2019aid,ALICE:2022dyy}. Surprisingly, the existing measurements for vector mesons $\phi$ and $J/\psi$ don't even have the common sign. Early theoretical studies based on coalescence model indicate that spin alignment occurs at quadratic order in the vorticity \cite{Yang:2017sdk,Goncalves:2021ziy}, which is too small to accommodate the magnitude of data. Different mechanisms of spin alignment have been explored \cite{Sheng:2019kmk,Sheng:2023urn,DeMoura:2023jzz,Kumar:2023ghs,Yang:2024qpy,Xu:2024kdh,Sheng:2024kgg,Zhu:2025rdj,Chen:2025mrf,Liang:2025hxw,Chen:2026ton,Yang:2024fkn}, but no converging picture has been reached. In this paper, we study a possible contribution induced by shear of QGP.

The tensorial nature of shear allows it to contribute to spin alignment at linear order \cite{Li:2022vmb,Wagner:2022gza,Dong:2023cng}. In fact, it can contribute to spin alignment through different mechanisms such as modifications to spectral functions \cite{Li:2022vmb,Dong:2023cng} and chromo field fluctuations \cite{Sheng:2019kmk,Liang:2026jna} etc. In this paper, we study another possible contribution through modification to vector meson structure. The picture is most transparent for quarkonia like $J/\psi$, which adopts a potential description. In an equilibrium QGP, the potential is Debye screened and can lead to suppression of $J/\psi$ \cite{Matsui:1986dk,Karsch:1987pv}. In an off-equilibrium QGP with shear, the potential is no longer spherical. The presence of shear modifies the distributions of medium quarks and gluons, which enter self-energy of exchanged gluon between the constituents of the quarkonia. We shall see that the shear contirbutes a tensor component of the potential, which further gives rise to a D-wave component of the wave function, see also related studies in the context of anisotropic QGP \cite{Dumitru:2009ni,Dong:2021gnb}. The D-wave component is precisely linked to the spin alignment \cite{Yan:2025tlx}.
While the mechanism in principle works for $\phi$ as well, it is not expected to impact significantly. This is because $J/\psi$ has suppressed hadrnoic interaction, but $\phi$ is strongly affected by hadrnoic interaction, so that the effect for $\phi$ may be easily washed out in hadronic phase. We shall restrict our study to quarkonia.

The paper is organized as follows: in Sec.~\ref{sec:potential}, we study the shear correction to the potential for the quarkonium. This is done by calculating the gluon self-energy in the hard thermal loop approximation, which leads to a tensor component of the potential. We further use perturbation theory to find the D-wave component of the wave function; in Sec.~\ref{sec:alignment}, we establish a relation between the D-wave component to the spin alignment from the angular distributin of leptonic decay of $J/\psi$. We also study phenomenology for $J/\psi$ spin alignment based on the mechanism in a Bjorken flow. We summarize and discuss future directions in Sec.~\ref{sec:summary}. Details of calculations are reserved to three appendices.

\section{The shear correction to the quarkonium potential}\label{sec:potential}
\subsection{The quarkonium and resummed gluon propagator}

We first express the quarkonium potential in terms of in-medium gluon propagator. The potential can be extracted from scattering of the quarkonium constituents. 
The coupling between the heavy quark and the gluon field is given by the vertex in the QCD Lagrangian:
\begin{align}
    \mathfrak{L} _{int}=g\bar{\psi}\gamma^\mu T^a A_\mu^a \psi
\end{align}
For heavy quarks with mass much larger than the QCD scale $\Lambda_{QCD}$ and the medium temperature $T$, their motion is non-relativistic with velocity $v\ll 1$. It follows that the time component of the color current is much larger than the spatial components $\bar{\psi}\gamma^0  \psi \gg \bar{\psi}\gamma^i  \psi$. Therefore, at leading order in the non-relativistic approximation, the spatial components can be neglected and the interaction between two static heavy quarks is dominated by the exchange of the temporal component of virtual gluons.
\begin{figure}[htbp]
    \centering
    \includegraphics[width=0.15\textwidth]{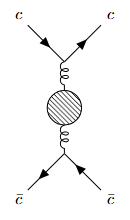}
    \caption{Schematic picture for charm and anti-charm scattering in medium, with blob denoting medium modification to exchangd gluon.}
    \label{fig:tree2}
\end{figure}
Performing non-relativistic expansion, we obtain the amplitude for the scattering of a quark-antiquark pair as
\begin{align}
    i\mathcal{M}&=-g^2\bar{u}\gamma_\mu T^au\bar{v}\gamma_{\nu}T^bv D_{F,ab}^{\mu\nu}\nonumber\\ &\simeq-4im^2g^2D_{F,ab}^{00}(Q)\xi^{\dagger}_iT_{ij}^a\xi_j\xi^{'\dagger}_{k}T_{kl}^b \xi'_{l},
\end{align}
where
\begin{align}
    u(p)\simeq\sqrt{m}
    \begin{pmatrix}
    \xi\\ 
    \xi
    \end{pmatrix},
    \qquad
  v(p)\simeq\sqrt{m}
        \begin{pmatrix}
        \xi' 
\\-\xi'
    \end{pmatrix}.  
\end{align}
Since the medium is color neutral, we have $D_{F,ab}\propto \delta_{ab}$, we have
\begin{align}
T^a_{ij}T^a_{kl}=\frac{1}{2}\left(\delta_{il}\delta_{jk}-\frac{1}{N_c}\delta_{ij}\delta_{kl}\right).
\end{align}
Performing color average for the singlet state $|1\rangle=\frac{1}{N_c^{1/2}}\sum_i|i\bar{i}\rangle$, we obtain the color factor $C_F=\frac{N_c^2-1}{2N_c}$ for singlet potential. Below we denote $D_{F,ab}=D_F\delta_{ab}$.

The medium effect encoded in $D_F^{00}$ contains both potential and damping \cite{Laine:2006ns,Carrington:2024ize}. The two effects can be separated as
\begin{align}
    D_{F}^{\mu\nu}=-iD_R^{\mu\nu}+D^{<\mu\nu},
\end{align}
with retarded and lesser propagator encoding potential and damping effects respectively. We shall focus on potential in this paper, thus we keep $D_R^{\mu\nu}$ below. 
\begin{align}
    i\mathcal{M} =-C_F{4im^2g^2}D_R^{00}(Q)\xi^\dagger\xi\xi'{}^\dagger\xi'.
\end{align}
In the non-relativistic limit, we have $q_0\simeq0$. We can extract the potential from the amplitude as
\begin{align}\label{potential}
    V(q)=g^2C_FD_R^{00}(q_0=0,q).
\end{align}
The full gluon propagator satisfies the Dyson equation:
\begin{align}\label{Dyson}
    D^{\mu \nu }=D_{0}^{\mu \nu}+D_{0}^{\mu \alpha} \Pi_{\alpha \beta}D^{\beta \nu}.
\end{align}
The propagators and self-energy are both retarded. We have suppressed the retarded index $R$ for notational simplicity.
We shall assume the quarkonium to be slow-moving with respect to the medium for simplicity, so that the non-relativistic expansion is valid in the QGP frame. In this case, only the longitudinal component of self-energy is relevant for equilibrium QGP, whose hard thermal loop (HTL) approximation tends to a constant in the static limit \footnote{For quarkonium moving with respect to QGP, both longitudinal and transverse components are relevant, see for example \cite{York:2008rr}.}:
\begin{align}\label{Debye}
    \Pi_{00}(\omega=0,q)=m_D ^2, \\
    m_D ^2=\frac{g^2 T^2}{3}(C_A+\frac{N_f}{2}),
\end{align}
where $C_A=N_c$ is the Casimir operator for $SU(N)$ and $N_f$ is the number of flavors \cite{Bellac:2011kqa}. In an off-equilibrium medium with shear, this screening mass is no longer a constant. The general structure of the self-energy up to first order in shear tensor can be written as
\begin{align}
    \Pi^{ \mu \nu}=\Pi_{T}P_{T}^{\mu\nu}+\Pi_{L}P_{L}^{\mu\nu}+\Pi_{TT}P_{TT}^{\mu\nu}+\Pi_{LT}P_{LT}^{\mu\nu}. \label{eq.(4)}
\end{align}
Here $P_T^{\mu\nu}$ and $P_L^{\mu\nu}$ are the transverse and longitudinal projectors defined as $P_{T}^{\mu\nu}=P^{\mu\nu}-\frac{P^{\mu\alpha}P^{\nu\beta}Q_{\alpha}Q_{\beta}}{q^2}$ with $P^{\mu\nu}=u^{\mu}u^{\nu}-\eta^{\mu\nu}$ and $P_{L}^{\mu\nu}=\frac{Q^\mu Q^\nu}{Q^2}-\eta^{\mu\nu}-P_T^{\mu\nu}$. $u^\mu$ is the fluid velocity. $\Pi_T$ and $\Pi_L$ can have first order shear correction.
The emergent projector $P_{TT}^{\mu\nu}$ and $P_{LT}^{\mu\nu}$ are constructed as \cite{Lin:2022tma}
\begin{align}
&P_{TT}^{\mu\nu}=P_{T}^{\mu\alpha}S_{\alpha \beta}P_{T}^{\beta\nu},\nonumber\\
&P_{LT}^{\mu\nu}=P_{L}^{\mu\alpha}S_{\alpha \beta}P_{T}^{\beta\nu}+P_{T}^{\mu\alpha}S_{\alpha \beta}P_{L}^{\beta\nu},\nonumber
\end{align}
Here $S^{\mu\nu}=\sigma^{\mu\nu}/T$ is the dimensionless shear tensor and the usual shear tensor $\sigma^{\mu\nu}$ is defined as
\begin{align}\label{shear}
    \sigma^{\mu\nu}=\frac{1}{2}P^{\mu\alpha}P^{\nu\beta}(\partial_{\alpha} u_{\beta}+\partial_{\beta} u_{\alpha})-\frac{1}{3}P^{\mu\nu}P^{\alpha\beta}\partial_\alpha u_{\beta}.
\end{align}
With $P_{TT}^{\mu\nu}$ and $P_{LT}^{\mu\nu}$ being first order in shear, $\Pi_{TT}$ and $\Pi_{LT}$ are zeroth order in shear.
Using \eqref{Dyson} and choose Coulomb gauge with $D_0^{\mu\nu}=\frac{-1}{Q^2}P_T^{\mu\nu}+\frac{-1}{q^2}u^\mu u^\nu$, we can obtain the resummed propagator up to first order in shear as 
\begin{align}
    D^{\mu\nu}=\frac{-1}{Q^2-\Pi_T}P_{T}^{\mu\nu}+\frac{-1}{q^2-\frac{Q^2}{q^2}\Pi_L}u^\mu u^\nu+\frac{-\Pi_{TT}}{(Q^2-\Pi_T)^2}P_{TT}^{\mu\nu}+\frac{-\Pi_{LT}}{(Q^2-\Pi_T)(Q^2-\Pi_L)}P_{LT}^{\mu\nu}. \label{eq.(5)}
\end{align}
Appendix A describes the properties of the general structure of the self-energy and the resummed propagator. Using the representation \eqref{potential}, which singles out only the longitudinal component of the resummed propagator, we find after Fourier transform
\begin{align}\label{V}
    V(r)=C_F \frac{g^2}{(2\pi)^3}\int d^3q \frac{1}{q^2+\Pi_{L}(q_0=0)}e^{-i\vec{q}\cdot\vec{r}}.
\end{align}
Therefore, we need the shear correction to longitudinal self-energy $\Pi_L$ in the static limit. We will obtain it in the HTL approximation.

\subsection{The shear correction to the gluon self-energy}

The retarded gluon self-energy can be conveniently described in the real time formalism of finite temperature field theory in the $ra$-basis, with the following relation
\begin{align}
    \Pi_{ar}^{\mu\nu}=-i\Pi_R^{\mu\nu}.
\end{align}
In an off-equilibrium medium with shear, the quarks and gluons can have polarized and unpolarized parts \cite{Lin:2026rkd}. The dominant effect is the shear correction to unpolarized distribution
\begin{align}
    f_{q/g}&=f^{(0)}_{q/g}+ f^{(1)}_{q/g} \nonumber\\
    f_{q/g}^{(1)}(p_0)&=\frac{\beta \sigma_{ij}}{2} \hat{q}_i \hat{q}_j \chi_{q/g} \nonumber\\
    \chi_{q/g}&=\beta^2p^2 \tilde{\chi}_{q/g} .
\end{align}
Here $f^{(0)}_{q/g}$ are the equilibrium distribution functions for quarks and gluons respectively. $f^{(1)}_{q/g}$ are the shear corrections. $sgn(p_0)$ is the sign function. $\beta=1/T$. $\chi_{q/g}$ are scalar function to be determined by solution to Boltzmann equation.
In terms of the modified distribution functions, the quark propagators in the $ra$-basis are given by
\begin{align}
    S_{ar(0)}(p)=\frac{i\slashed{P}}{(p_0-i\epsilon)^2-p^2}, \nonumber\\
    S_{ra(0)}(p)=\frac{i\slashed{P}}{(p_0+i\epsilon)^2-p^2}, \nonumber\\
    S_{rr(0)}(p)=(\frac{1}{2}-f_q(p_0))2\pi sgn(p_0)\slashed{P} \delta(P^2)
\end{align}
and the counetrparts for gluons in Coulomb gauge are given by
\begin{align}
    D_{ar(0)}^{\mu\nu}(Q)=\frac{i}{(q_0-i\epsilon)^2-q^2}(P_T^{\mu\nu}+\frac{Q^2}{q^2}u^\mu u^\nu),  \nonumber\\
    D_{ra(0)}^{\mu\nu}(Q)=\frac{i}{(q_0+i\epsilon)^2-q^2}(P_T^{\mu\nu}+\frac{Q^2}{q^2}u^\mu u^\nu),  \nonumber\\
    D_{rr(0)}^{\mu\nu}(Q)=2\pi sgn(q_0) \delta(Q^2)(\frac{1}{2}-f_g(q_0))(P_T^{\mu\nu}+\frac{Q^2}{q^2}u^\mu u^\nu)
\end{align}
We have suppressed the trivial color factors in the above for notational simplicity.

\begin{figure}[H]
    \centering
    \includegraphics[width=1\textwidth]{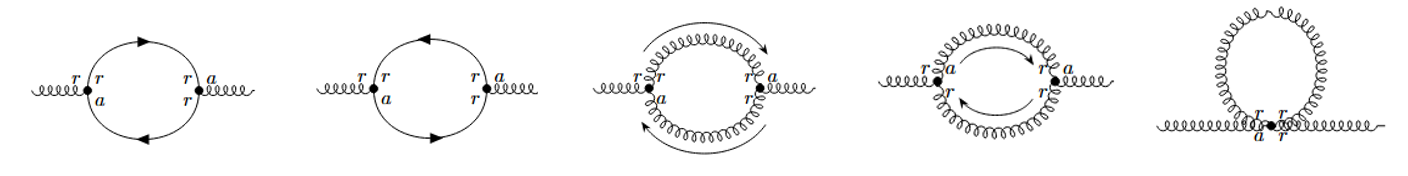}
    \caption{The gluon self-energy diagram}
    \label{fig:HTL}
\end{figure}

The gluon self-energy in the $ra$-basis can be obtained in the HTL approximation \cite{York:2008rr}
\begin{align}\label{Debye_HTL}
    \Pi^{00(1)}_{ar}&=-i\beta \delta m^2 A^{00}, \nonumber\\
    A^{00}&=\frac{q_i q_j \sigma_{ij}}{2q^2}[2(3x^2-1)-(3x^3-2x)ln(\frac{x+1+i\epsilon}{x-1+i\epsilon})], \nonumber\\
    \delta m^2&=N_f \frac{g^2\beta^2}{\pi^2} \int dp p^3 \tilde{\chi}_q+C_A \frac{g^2\beta^2}{\pi^2} \int dp p^3 \tilde{\chi}_g,
\end{align}
where the parameter $x$ is defined as $x=\frac{q_0}{q}$. 

Unlike in equilibrium, the self-energy in sheared QGP depends on the direction of momentum through $q_iq_j\sigma_{ij}$, which will break the spherical symmetry of the quarkonium potential. At $x=0$, $\Pi_L=-\Pi^{00}$.

For phenomenological application, we adopt the following solutions of $\tilde{\chi}_{q/g}$ \cite{Dusling:2009df}
\begin{align}
        \tilde{\chi}_{q}&=-7.75sgn(p_0) \beta^{-\alpha_q} p^{-\alpha_q} \frac{\eta}{s} f_q^{(0)}(1-f_q^{(0)}) \nonumber\\
    \tilde{\chi}_{g}&=-4.5sgn(p_0) \beta^{-\alpha_g} p^{-\alpha_g} \frac{\eta}{s} f_g^{(0)}(1+f_g^{(0)}).
\end{align}
with the exponents $\alpha_q \approx \alpha_g \approx 0.38$ and phenomenological value $\eta/s=0.08$ is used. Appendix B shows the evolution of gluon self-energy for shear. The $00$ component of the self-energy is reduced as
\begin{align}
    \Pi_{ar}^{00,AB(1)}(q_0=0)=-(19.38N_f+18.25C_A)\frac{\eta}{s} \delta^{AB} \frac{ig^2}{\pi^2\beta} \frac{q_iq_j\sigma_{ij}}{q^2}.
\end{align}
Expanding \eqref{V} to first order in the shear tensor, we find the perturbation to potential from shear in coordinate space as
\begin{align}
    \triangle V&=\frac{r_ir_j\sigma_{ij}}{r^2}f(r), \nonumber\\
    f(r)&=-C_F\frac{g^2\beta \delta m^2}{4\pi}(\frac{e^{-m_D r}}{2m_D}+\frac{3(e^{-m_D}-1)}{m_D^4 r^3}+\frac{3e^{-m_D r}}{m_D^3 r^2}+\frac{3e^{-m_D r}}{2m_D^2 r}).
\end{align}
The Appendix C shows how obtain the potential for shear in coordinate space by Fourier transform. 

The vector meson $J/\psi$ corresponds to the spin triplet S-wave state of charmonium $c\bar{c}$. The shear tensor introduces a D-wave compoennt to the potential, which further gives rise to D-wave component of the $J/\psi$ wave function. The D-wave component can be obtained by the perturbation theory as
\begin{align}
    \psi_D(q)&=\sum_{n}\frac{\left \langle n20 \left | \triangle  V  \right | 100 \right \rangle }{E_{n20}-E_{100}}\left \langle q |  n20  \right \rangle   \nonumber\\
    &=Y_2^0(\Omega_q)R_D(q)  \nonumber\\
    &=Y_2^0(\Omega_q) \int Y_2^0(\Omega_r) \frac{r_ir_j\sigma_{ij}}{r^2} Y_0^0(\Omega_r) d\Omega_r \tilde{R}_D(q) \nonumber\\
    \tilde{R}_D(q)&=\sum_n \frac{\int r^2 R_{10}(r)f(r)R_{n2}(r)dr}{E_{n20}-E_{100}}R_{n2}(q).
\end{align}
Here, $E_{n20}$ and $E_{100}$ are energies of $\left | n20  \right \rangle  $ and $\left | 100  \right \rangle  $ respectively. $Y_l^m(\Omega_q)$ is spherical harmonics and $R_{nl}(q)$ is radial wave function.

\section{The quarkonium spin alignment}\label{sec:alignment}
\subsection{The angular distribution of vector meson decay}
The spin alignment of vector mesons is determined by measuring the diagonal element $\rho_{00}$ of the spin density matrix in experiments. $\rho_{00}$ represents the probability of spin polarization of the vector meson in the state with spin magnetic quantum number $m=0$. If there is no spin alignment, its value should be $\frac{1}{3}$, which means that the three spin projection states of the vector meson have the same probability. $\rho_{00}$ is extracted indirectly by analyzing the angular distribution of its decay products
\begin{align}
    \frac{dN}{dcos\theta} \propto (1-\rho_{00})+(3\rho_{00}-1)cos^2\theta.
\end{align}
For the vector meson $J/\psi$, its decay products are a lepton-antilepton pair $l\bar{l}$. The decay width is given by

\begin{figure}[htbp]
    \centering
    \includegraphics[width=0.3\textwidth]{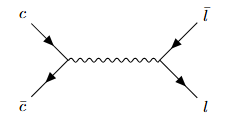}
    \caption{$J/\psi\longrightarrow l\bar{l}$.}
    \label{fig:tree}
\end{figure}

\begin{align}\label{angular}
    d\Gamma&=\frac{e^4}{Q^4}\int \frac{d^3p_1}{2E_1(2\pi)^3} \frac{d^3p_2}{2E_2(2\pi)^3} (2\pi)^4 \delta^4(Q-P_1-P_2) l_{\mu\nu} \Pi^{\mu\nu}_{photon}.
\end{align}
Here $p_1$ and $p_2$ is momenta of lepton and anti-lepton and $l_{\mu\nu}=4[p_{1\mu}p_{2\nu}+p_{2\mu}p_{1\nu}-(p_1 \cdot p_2)g_{\mu\nu}]$. $\Pi_{photon}^{\mu\nu}$ is the self-energy of the intermediate photon in the decay process of vector meson $J/\psi$. It satisfies the Ward identity $Q_\mu\Pi_{photon}^{\mu\nu}=0$. In the photon rest frame, we have:
\begin{align}
    \Pi_{photon}^{00}=\Pi_{photon}^{0i}=\Pi_{photon}^{i0}=0,
\end{align}
By our assumption, the photon rest frame coincides with the QGP frame. In the absence of shear, the photon self-energy remains invariant under rotation $\Pi_{photon}^{ij}  \propto \delta^{ij}$. The shear tensor introduces a D-wave component to the quarkonium state, which is inherited by the virtual photon state. The photon self-energy can be decomposed into scalar and tensor parts
\begin{align}\label{photon_Pi}
    \Pi_{photon}^{ij}=\Pi_S \delta^{ij}+\Pi_T \sigma^{ij}.
\end{align}
Plugging \eqref{photon_Pi} into \eqref{angular} and performing integration over azimuthal angle, we obtain the following decay angular distribution
\begin{align}
    \frac{d\Gamma}{dcos\theta}&=\frac{e^4}{4\pi M^4}(4\Pi_s+\Pi_Tl_il_j\sigma_{ij}-3cos^2\theta l_il_j\sigma_{ij}\Pi_T),
\end{align}
where $M$ is the mass of $J/\psi$. The diagonal element of the spin density matrix $\rho_{00}$ can be obtained from the decay angular distribution
\begin{align}
    \rho_{00}&=\frac{1}{3}+\frac{l_i l_j\sigma_{ij}\Pi_T}{3\Pi_S}
\end{align}
where $l_i$ is the unit vector along the quantization axis. The D-wave component in the wave function gives rise to the following photon self-energy
\begin{align}
    \Pi^{ij}_{photon} &\propto \delta^{kl} \int_{q,k} (\delta^{ik}+\alpha_q \hat{q}^i \hat{q}^k)(\delta^{jl}+\alpha_k \hat{k}^j \hat{k}^l) \psi(q) \psi^*(k) \nonumber\\
    &=\int_{q,k} (\delta^{ij}+\alpha_q \hat{q}^i \hat{q}^j+\alpha_k \hat{k}^i \hat{k}^j+\alpha_q \alpha_k \hat{q}^i \hat{k}^j \hat{q}\cdot \hat{k})(\psi_0(q)+\psi_D(q))(\psi_0(k)+\psi_D(k)), \label{photon_se}
\end{align}
where $\psi_0=Y_0^0(\Omega_q)R_0(q)$ and $\alpha_q=-\frac{q^2}{E_q(m+E_q)}$ \cite{Yan:2025tlx}. Using the following relations:
\begin{align}
    &\int  Y_0^0(\Omega_q) \delta_{ij} d\Omega_q=\sqrt{4\pi}\delta_{ij}, \nonumber\\
    &\int   Y_2^0(\Omega_q) \delta_{ij} d\Omega_q=0, \nonumber\\
    &\int  Y_0^0(\Omega_q) \frac{q_iq_j \sigma_{ij}}{q^2} d\Omega_q=\frac{\sqrt{4\pi}}{3}\delta_{ij}, \nonumber\\
    &\int Y_2^0(\Omega_q) \frac{q_iq_j \sigma_{ij}}{q^2} d\Omega_q=\frac{4\sqrt{\pi}}{135}\sigma_{ij},
\end{align}
in performing the integration of angular variable $\Omega_q$, we can obtain from eq.\eqref{photon_se} up to the first order in the D-wave component
\begin{align}
     \Pi^{ij}_{photon} &\propto \int q^2[(1+\frac{\alpha_q}{3})R_0(q)\delta^{ij}+\frac{4}{135}\alpha_q \tilde{R}_D(q)\sigma^{ij}] dq.
\end{align}
This allows us to express $\Pi_S$ and $\Pi_T$ in terms of S-wave and D-wave components of wave function as
\begin{align}
    \Pi_S&=\int q^2(1+\frac{\alpha_q}{3})R_0(q) dq, \nonumber\\
    \Pi_T&=\int \frac{4}{135}\alpha_q q^2\tilde{R}_D(q) dq.
\end{align}
The diagonal element of the spin density matrix is given by
\begin{align}
    \rho_{00}=\frac{1}{3}+\frac{4}{405\tau}\frac{\int \frac{4}{135}\alpha_q q^2\tilde{R}_D(q) dq}{\int q^2(1+\frac{\alpha_q}{3})R_0(q) dq}.
\end{align}

\subsection{Shear induced spin alignment}

In this section, we study the shear induced contribution to spin alignment.
In this paper, we use the Bjorken model for fluid velocity. The Bjorken flow model assumes that the collision is head-on and that the matter is uniformly distributed in the transverse plane perpendicular to the beam direction, so there is no transverse expansion, only one-dimensional longitudinal expansion. The fluid velocity of Bjorken model in lab frame is given by
\begin{align}
    u_\mu=(\frac{t}{\tau},0,0,\frac{z}{\tau})
\end{align}
where $\tau=\sqrt{t^2-z^2}$ is the proper time and $t$ is time. Substituting the flow velocity of Bjorken model in lab frame into \eqref{shear}, the individual components of the shear tensor in lab frame can be obtained as
\begin{align}
    \sigma^{\mu\nu}_{lab}=
    \begin{pmatrix}
  -\frac{2z^2}{3\tau^3}& 0 & 0 & -\frac{2zt}{3\tau^3} \\
 0 &\frac{1}{3\tau}  & 0 & 0\\
 0 & 0 & \frac{1}{3\tau} &0 \\
 -\frac{2zt}{3\tau^3} & 0 & 0 & -\frac{2t^2}{3\tau^3}
\end{pmatrix}.
\end{align}
We shall perform Lorentz boost to the fluid frame, in which the gluon self-energy is evaluated. The shear tensor in the fluid frame where $u_\mu=(1,0,0,0)$ reads
\begin{align}
    \sigma^{\mu\nu}_{fluid}=
    \begin{pmatrix}
 0 & 0 & 0 & 0\\
 0 & \frac{1}{3\tau} & 0 & 0\\
 0 & 0 & \frac{1}{3\tau} & 0\\
 0 & 0 & 0 &-\frac{2}{3\tau}
\end{pmatrix}.
\end{align}
The quantization axis is perpendicular to the event plane, thus $l^\mu=(0,0,1,0)$ in the lab frame. It does not change under longitudinal boost from lab frame to fluid frame. Thus we have $l_jl_j\sigma_{ij}=\frac{1}{3\tau}$.
We choose the parameters proper time $\mu$ to be $\tau=10 fm$ and the reduced mass $\mu=0.65GeV$, the corresponding $\rho_{00}-\frac{1}{3}$ is obtained in the illustrative range of the Debye mass $m_D^2$.
\begin{figure}[H]
    \centering
    \includegraphics[width=0.8\textwidth]{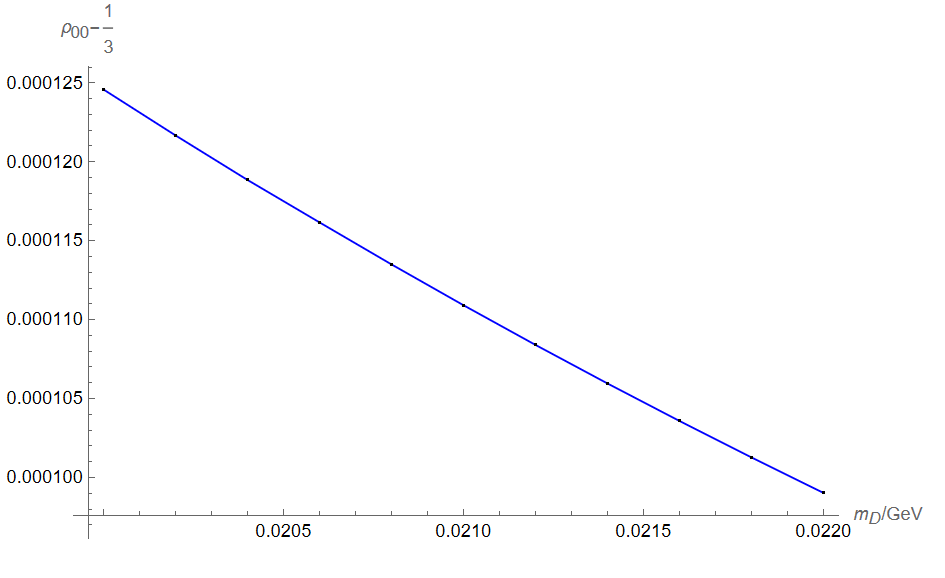}\caption{
     The value of $\rho_{00}-\frac{1}{3}$ for proper time $\tau=10 fm$ and reduced mass $\mu=0.65GeV$ }
    \label{fig:spin}
\end{figure}
When the actual Debye mass from \eqref{Debye} at critical temperature and realistic coupling are used, the D-wave of the quarkonium dissolves into the continuum, invalidating \eqref{photon_se} based on discrete spectrum. In fact, the survival of D-wave quarkonium is sensitive to the potential function between heavy quarks. For long range Coulomb interaction used above, the seperation of heavy quark–antiquark pair is far, so that their interaction is easily screened. Realistic interaction also includes a short range string term, making the quarkonium more compact and harder to dissolve. For the potential function in a finite-temperature QCD plasma, we can use the Karsch-Mehr-Satz (KMS) model \cite{Dumitru:2009ni}:
\begin{align}
    V_{KMS}(r)=-\frac{\alpha}{r}(1+m_D r)exp(-m_D r)+2 \frac{\sigma}{m_D}[1-exp(-m_D r)]-\sigma r exp(-m_D r).
\end{align}
Here, $\alpha$ is an effective Coulomb coupling at short distances and $\sigma$ is the string tension. By incorporating the entropy contribution, this potential function generates an overall deeper and more attractive potential. It predicts a larger binding energy and a higher dissociation temperature, indicating that quarkonium can survive in the plasma up to higher temperatures than the screened Coulomb potential used before. At the phase transition temperature $T_c$, a weakly bound D-wave state exists. Following the treatment in \cite{Dumitru:2009ni}, we use the shear corrected anisotropic Debye mass obtained by HTL \eqref{Debye_HTL} in place of isotropic Debye mass, the perturbative potential function from shear expanded to first order can be obtained as
\begin{align}
    \triangle V_{KMS}(r)&=\frac{\sigma_{ij}r_i r_j}{r^2} f_{KMS}(r)     ,\nonumber\\
    f_{KMS}(r)&=2\pi \beta \delta m^2 [-\alpha \frac{A}{m^4_D r^3}+(\alpha m_D^2-\sigma)\frac{B}{m^6_D r^3}+\sigma \frac{C}{m_D^7 r^4}] ,\nonumber\\
    A&=6-e^{-m_D r}(m_D^3 r^3 +3m_D^2 r^2+6m_D r+6), \nonumber\\
    B&=24-e^{-m_D r}(m_D^4 r^4+4m_D^3 r^3+12m_D^2 r^2+24m_D r+24), \nonumber\\
    C&=144-e^{-m_D r}(m_D^5 r^5+6m_D^4 r^4+24m_D^3 r^3+72m_D^2 r^2+144m_D r+144).
\end{align}
Appendix D collects the derivation of the perturbative potential function from shear expanded to first order using the anisotropic Debye mass. Using the same method as for the Yukawa potential, the spin alignment value of the KMS model at the phase transition temperature $T_c=157MeV$  can be obtained
\begin{align}
    \rho_{00}-\frac{1}{3}=-0.00133.
\end{align}
Here, the Debye mass at the phase transition temperature $T_c$ is $m_D=0.435GeV$. The parameters $\alpha$ and $\sigma$ are chosen to be $\alpha=0.385$ and $\sigma=0.223GeV^2$.

\section{Summary}\label{sec:summary}
In this paper, we study shear induced spin alignment for quarkonium. With the gluon gaining an anisotropic Debye mass in QGP with shear flow, the quarkonium potential receives a tensor type correction, which gives rise to a D-wave component of the quarkonium wave function. The D-wave component leads to an orbital contribution to spin alignment. The shear correction to Debye mass is calculated in the HTL approximation. Application of the mechanism to $J/\psi$ spin alignment in heavy ion collisions has been discussed. D-wave charmonium doesn't survive at QGP temperature when screened Coulomb potential is used. When realistic screened potential is used, the D-wave charmonium can exist at critical temperature and $J/\psi$ receives a D-wave component in QGP with shear. The orbital contribution to spin alignment is found to be negative with a magnitude of $10^{-3}$.

For simplicity, we have restricted ourselves to quarkonium at rest in QGP. In this case, only electric gluons mediate the interaction between heavy quark-antiquark pair. For quarkonium moving with respect to QGP, magnetic gluons also contribute, which is expected to generate shear induced spin dependent potential. This will be complementary to orbital contribution to spin alignment. We leave this for future studies.

\section*{Acknowledgments}
We thank Yun Guo and Defu Hou for stimulating discussions. This work is in part supported by NSFC under Grant Nos 12475148, 12075328.

\appendix

\section{The resummed propagator}

In this section, we show how to obtain the self-energy and the resummed propagator. We start from the self-energy for shear
\begin{align}
    \Pi^{ \mu \nu}=\Pi_{T}P_{T}^{\mu\nu}+\Pi_{L}P_{L}^{\mu\nu}+\Pi_{TT}P_{TT}^{\mu\nu}+\Pi_{LT}P_{LT}^{\mu\nu}  
\end{align}
where $S^{\mu\nu}=\sigma^{\mu\nu}/T$ is the dimensionless shear tensor and the dimension of $S_{\mu\nu}$ is $S_{\mu\nu}\sim O(\partial)$. The bare propagator is
\begin{align}
    D_0^{\mu\nu}=-\frac{1}{Q^2}P_T^{\mu\nu}-\frac{1}{q^2}u^\mu u^\nu
\end{align}
where we choose Coulomb gauge. We expand in powers of $\partial$ and keep it up to the first-order of $\partial$. The resummed propagator for shear can be obtained as
\begin{align}
    D^{\mu\nu}=-\frac{1}{Q^2-\Pi_T}P_{T}^{\mu\nu}-\frac{1}{q^2-\frac{Q^2}{q^2}\Pi_L}u^\mu u^\nu-\frac{\Pi_{TT}}{(Q^2-\Pi_T)^2}P_{TT}^{\mu\nu}-\frac{\Pi_{LT}}{(Q^2-\Pi_T)(Q^2-\Pi_L)}P_{LT}^{\mu\nu}. 
\end{align}
The self-energy for shear can be reduced as
\begin{align}    \Pi^{\mu\nu}&=\Pi_{T}P_{T}^{\mu\nu}+\Pi_{L}P_{L}^{\mu\nu}+\Pi_{TT}\beta[\sigma^{\mu\nu}-\frac{\omega}{q^2}a^{(\mu}u^{\nu)}+\frac{1}{q^2}a^{(\mu}Q^{\nu)}+\frac{\omega^2 Q\cdot a}{q^4}u^\mu u^\nu-\frac{\omega Q\cdot a}{q^4}u^{(\mu}Q^{\nu)}+\frac{Q \cdot a}{q^4}Q^\mu Q^\nu] \nonumber\\
    &+\Pi_{LT}\beta[\frac{\omega}{q^2} a^{(\mu}u^{\nu)}-\frac{1}{q^2}a^{(\mu}Q^{\nu)}-\frac{\omega^2 Q\cdot a}{q^4}u^\mu u^\nu+\frac{\omega Q\cdot a}{q^4}u^{(\mu}Q^{\nu)}-\frac{Q \cdot a}{q^4}Q^\mu Q^\nu] 
\end{align}
where $a^\mu=\sigma^{\mu\nu}Q_\nu$.The symbol $A^{(\mu}B^{\nu)}$ is defined as 
\begin{align}
    A^{(\mu}B^{\nu)}=A^{\mu}B^{\nu}+A^{\nu}B^{\mu}. 
\end{align}
In the fluid frame, $P_T^{00}=0$ and $\sigma^{00}=0$. The longitudinal projector in Coulomb gauge is $P_L^{00}=-1$. The self energy for shear in 00 component is
\begin{align}
    \Pi^{00}=-\Pi_{L}. 
\end{align}
The resummed propagator for shear can be reduced as
\begin{align}
    D^{\mu\nu}&=-\frac{1}{Q^2-\Pi_T}P_T^{\mu\nu} -\frac{1}{q^2-\frac{Q^2}{q^2}\Pi_L}u^\mu u^\nu \nonumber\\
    &-\frac{\Pi_{TT}}{(Q^2-\Pi_T)^2}\beta[\sigma^{\mu\nu}-\frac{\omega}{q^2}a^{(\mu}u^{\nu)} +\frac{1}{q^2}a^{(\mu}Q^{\nu)}+\frac{\omega^2 Q\cdot a}{q^4}u^\mu u^\nu-\frac{\omega Q\cdot a}{q^4}u^{(\mu}Q^{\nu)}+\frac{Q \cdot a}{q^4}Q^\mu Q^\nu] \nonumber\\
    &-\frac{\Pi_{LT}}{(Q^2-\Pi_T)(Q^2-\Pi_L)}\beta[\frac{\omega}{q^2}a^{(\mu}u^{\nu)} -\frac{1}{q^2}a^{(\mu}Q^{\nu)}-\frac{\omega^2 Q\cdot a}{q^4}u^\mu u^\nu+\frac{\omega Q\cdot a}{q^4}u^{(\mu}Q^{\nu)}-\frac{Q \cdot a}{q^4}Q^\mu Q^\nu]+O(\partial^2). 
\end{align}
In the 00 component, the resummed propagator is
\begin{align}
    D^{00}=-\frac{1}{q^2-\frac{Q^2}{q^2}\Pi_L}. 
\end{align}

\section{The self-energy for shear}
In this section, we show the evolution of gluon self-energy for shear. Firstly, we start from the quark loop.

\begin{figure}[H]
    \centering
    \includegraphics[width=0.5\textwidth]{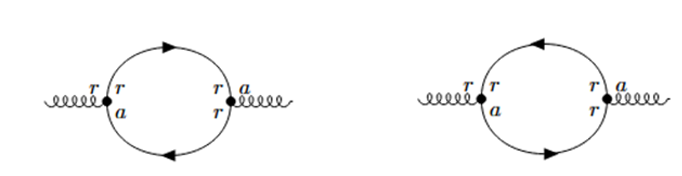}
    \label{fig:HTL1}
\end{figure}

\begin{align}
    \Pi_{ar}^{\mu\nu,AB}&=N_f g^2 \int \frac{d^4p}{(2\pi)^4}Tr[\gamma^\mu t_A S_ar(P-Q)\gamma^\nu t_B S_{rr}(P)]+N_f g^2 \int \frac{d^4p}{(2\pi)^4}Tr[\gamma^\mu t_A S_ra(P+Q)\gamma^\nu t_B S_{rr}(P)] \nonumber\\
    &=N_f \frac{1}{2} \delta^{AB} ig^2 \int \frac{d^4p}{(2\pi)^4} \frac{4((P-Q)^\mu P^\nu -g^{\mu\nu}(P-Q)\cdot P+P^\mu (P-Q)^\nu)}{(p_0-q_0-i\epsilon)^2-(\vec{p}-\vec{q})^2}(\frac{1}{2}-f_q)2\pi \cdot sgn(p_0) \delta(P^2) \nonumber\\
    &+N_f \frac{1}{2} \delta^{AB} ig^2 \int \frac{d^4p}{(2\pi)^4} \frac{4((P+Q)^\mu P^\nu -g^{\mu\nu}(P+Q)\cdot P+P^\mu (P+Q)^\nu)}{(p_0+q_0+i\epsilon)^2-(\vec{p}+\vec{q})^2}(\frac{1}{2}-f_q)2\pi \cdot sgn(p_0) \delta(P^2)
\end{align}
We evolve the self-energy with hard thermal loop $T \sim p \gg q$ and take $q$ parallel to the Z-axis, So $\frac{1}{(p_0-q_0-i\epsilon)^2-(\vec{p}-\vec{q})^2} \sim -\frac{1}{2}\frac{1}{p_0q_0-pqcos\theta+i\epsilon p_0-\frac{Q^2}{2}}$ and $\frac{1}{(p_0+q_0+i\epsilon)^2-(\vec{p}+\vec{q})^2} \sim \frac{1}{2}\frac{1}{p_0q_0-pqcos\theta+i\epsilon p_0+\frac{Q^2}{2}}$ . 

In 00 component, the self-energy is
\begin{align}
    \Pi_{ar}^{00,AB}&=-N_f \frac{1}{2} \delta^{AB} ig^2 \int \frac{d^4p}{(2\pi)^3} \frac{2p^2-p_0q_0-pqcos\theta}{p_0q_0-pqcos\theta+i\epsilon p_0-\frac{Q^2}{2}}(\frac{1}{2}-f_q)\frac{1}{p}(\delta(p_0-p)+\delta(p_0-p)) \nonumber\\
    &+N_f \frac{1}{2} \delta^{AB} ig^2 \int \frac{d^4p}{(2\pi)^3} \frac{2p^2+p_0q_0+pqcos\theta}{p_0q_0-pqcos\theta+i\epsilon p_0+\frac{Q^2}{2}}(\frac{1}{2}-f_q)\frac{1}{p}(\delta(p_0-p)+\delta(p_0-p))  \nonumber\\
    &=-N_f \frac{1}{2} \delta^{AB} ig^2 \int \frac{d^3p}{(2\pi)^3} ( \frac{2p-q_0-qcos\theta}{pq_0-pqcos\theta+i\epsilon p-\frac{Q^2}{2}}-\frac{2p+q_0-qcos\theta}{pq_0+pqcos\theta+i\epsilon p-\frac{Q^2}{2}})(\frac{1}{2}-f_q) \nonumber\\
    &+N_f \frac{1}{2} \delta^{AB} ig^2 \int \frac{d^3p}{(2\pi)^3} ( \frac{2p+q_0+qcos\theta}{pq_0-pqcos\theta+i\epsilon p+\frac{Q^2}{2}}-\frac{2p-q_0+qcos\theta}{pq_0+pqcos\theta+i\epsilon p-\frac{Q^2}{2}})(\frac{1}{2}-f_q).
\end{align}
The self-energy in the first order of $\partial$ is
\begin{align}
    \Pi^{00,AB(1)}_{ar}&=-N_f \frac{1}{2} \delta^{AB} ig^2 \int \frac{d^3p}{(2\pi)^3} ( \frac{2p-q_0-qcos\theta}{pq_0-pqcos\theta+i\epsilon p-\frac{Q^2}{2}}-\frac{2p+q_0-qcos\theta}{pq_0+pqcos\theta+i\epsilon p-\frac{Q^2}{2}})f_q^{(1)} \nonumber\\
    &+N_f \frac{1}{2} \delta^{AB} ig^2 \int \frac{d^3p}{(2\pi)^3} ( \frac{2p+q_0+qcos\theta}{pq_0-pqcos\theta+i\epsilon p+\frac{Q^2}{2}}-\frac{2p-q_0+qcos\theta}{pq_0+pqcos\theta+i\epsilon p-\frac{Q^2}{2}})f_q^{(1)}  \nonumber\\
    &=-N_f \frac{\beta^3 q_iq_j\sigma_{ij}}{8q^2} \delta^{AB} ig^2 \int \frac{d^3p}{(2\pi)^3} (\frac{(2p-q_0-qcos\theta)(3cos^2\theta-1)p^2}{pq_0-pqcos\theta+i\epsilon p-\frac{Q^2}{2}}-\frac{(2p+q_0-qcos\theta)(3cos^2\theta-1)p^2}{pq_0+pqcos\theta+i\epsilon p-\frac{Q^2}{2}})\tilde{\chi}_q  \nonumber\\
    &+N_f \frac{\beta^3 q_iq_j\sigma_{ij}}{8q^2} \delta^{AB} ig^2 \int \frac{d^3p}{(2\pi)^3} (\frac{(2p+q_0+qcos\theta)(3cos^2\theta-1)p^2}{pq_0-pqcos\theta+i\epsilon p+\frac{Q^2}{2}}-\frac{(2p-q_0+qcos\theta)(3cos^2\theta-1)p^2}{pq_0+pqcos\theta+i\epsilon p-\frac{Q^2}{2}})\tilde{\chi}_q
\end{align}
where the relationship we use is $p_ip_j \sigma_{ij}=p^2(3cos^2\theta-1)\frac{q_iq_j\sigma_{ij}}{2q^2}$. Then, we expand it by the order of $\frac{1}{p}$
\begin{align}
    \frac{(2p-q_0-qcos\theta)p^2}{pq_0-pqcos\theta+i\epsilon p-\frac{Q^2}{2}}=\frac{2p^2}{q_0-qcos\theta+i\epsilon}+\frac{q^2(cos^2\theta-1)}{(q_0-qcos\theta+i\epsilon)^2}p+O((\frac{1}{p})^0), \nonumber\\
    \frac{(2p+q_0-qcos\theta)p^2}{pq_0+pqcos\theta+i\epsilon p+\frac{Q^2}{2}}=\frac{2p^2}{q_0+qcos\theta+i\epsilon}-\frac{q^2(cos^2\theta-1)}{(q_0+qcos\theta+i\epsilon)^2}p+O((\frac{1}{p})^0), \nonumber\\
    \frac{(2p+q_0+qcos\theta)p^2}{pq_0-pqcos\theta+i\epsilon p+\frac{Q^2}{2}}=\frac{2p^2}{q_0-qcos\theta+i\epsilon}-\frac{q^2(cos^2\theta-1)}{(q_0-qcos\theta+i\epsilon)^2}p+O((\frac{1}{p})^0), \nonumber\\
    \frac{(2p-q_0+qcos\theta)p^2}{pq_0+pqcos\theta+i\epsilon p-\frac{Q^2}{2}}=\frac{2p^2}{q_0+qcos\theta+i\epsilon}+\frac{q^2(cos^2\theta-1)}{(q_0+qcos\theta+i\epsilon)^2}p+O((\frac{1}{p})^0). \nonumber\\
\end{align}
The first term in the order of $p^2 \sim T^2$ is
\begin{align}
    \Pi^{00,AB(1)}_{ar}&=-N_f \frac{\beta^3 q_iq_j\sigma_{ij}}{8q^2} \delta^{AB} ig^2 \int \frac{d^3p}{(2\pi)^3}(\frac{2p^2(3cos^2-1)}{q_0-qcos\theta+i\epsilon}-\frac{2p^2(3cos^2-1)}{q_0+qcos\theta+i\epsilon})\tilde{\chi}_q \nonumber\\
    &+N_f \frac{\beta^3 q_iq_j\sigma_{ij}}{8q^2} \delta^{AB} ig^2 \int \frac{d^3p}{(2\pi)^3}(\frac{2p^2(3cos^2-1)}{q_0-qcos\theta+i\epsilon}-\frac{2p^2(3cos^2-1)}{q_0+qcos\theta+i\epsilon})\tilde{\chi}_q \nonumber\\
    &=0.
\end{align}
Here, particles and antiparticles cancel each other. The second term  in the order of $p \sim T$ is
\begin{align}
    \Pi^{00,AB(1)}_{ar}&=-N_f \frac{\beta^3 q_iq_j\sigma_{ij}}{8q^2} \delta^{AB} ig^2 \int \frac{d^3p}{(2\pi)^3}(\frac{q^2(cos^2\theta-1)(3cos^2\theta-1)}{(q_0-qcos\theta+i\epsilon)^2}+\frac{q^2(cos^2\theta-1)(3cos^2\theta-1)}{(q_0+qcos\theta+i\epsilon)^2})p\tilde{\chi}_q \nonumber\\
    &+N_f \frac{\beta^3 q_iq_j\sigma_{ij}}{8q^2} \delta^{AB} ig^2 \int \frac{d^3p}{(2\pi)^3}(-\frac{q^2(cos^2\theta-1)(3cos^2\theta-1)}{(q_0-qcos\theta+i\epsilon)^2}-\frac{q^2(cos^2\theta-1)(3cos^2\theta-1)}{(q_0+qcos\theta+i\epsilon)^2})p\tilde{\chi}_q \nonumber\\
    &=N_f \frac{\beta^3 q_iq_j\sigma_{ij}}{2\pi^2q^2}  \delta^{AB} ig^2 \int dp p^3\tilde{\chi}_q[2(3x^2-1)-(3x^3-2x)ln(\frac{q_0+q+i\epsilon}{q_0-q+i\epsilon})
\end{align}
The gluon loop diagrams:
\begin{figure}[htbp]
    \centering
    \includegraphics[width=0.3\textwidth]{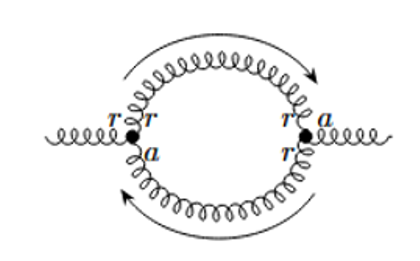}
    \label{fig:HTL2}
\end{figure}

The self-energy is
\begin{align}
    \Pi_{ar}^{\mu\nu,AB}&=\frac{1}{2}g^2f_{ADC}f_{BEF}\delta^{DE}\delta^{CF} \int \frac{d^4K}{(2\pi)^4}D_{ra}^{\alpha\beta}(Q+K)D_{rr}^{\sigma\rho}(K)[g_{\mu\alpha}(2Q+K)_\rho+g_{\alpha\rho}(-Q-2K)_\mu+g_{\rho\mu}(K-Q)_\alpha] \nonumber\\
    &\times[-g_{\nu\beta}(2Q+K)_\sigma+g_{\beta\sigma}(Q+2K)_\nu+g_{\sigma\nu}(Q-K)_\beta] \nonumber\\
    &=-\frac{1}{2}C_Aig^2 \int \frac{d^4K}{(2\pi)^4}\epsilon(k_0)\delta(K^2)(\frac{1}{2}+f_g)\frac{N^{\mu\nu}}{(q_0+k_0-i\epsilon)^2-(q+k)^2} \nonumber\\
    N_{\mu\nu}&=(P_T^{\alpha\beta}(Q+K)+\frac{(Q+K)^2}{(q+k)^2}u^\alpha u^\beta)[g_{\mu\alpha}(2Q+K)_\rho+g_{\alpha\rho}(-Q-2K)_\mu+g_{\rho\mu}(K-Q)_\alpha] \nonumber\\
    &\times(P_T^{\sigma\rho}(K)+\frac{K^2}{k^2}u^\sigma u^\rho)[-g_{\nu\beta}(2Q+K)_\sigma+g_{\beta\sigma}(Q+2K)_\nu+g_{\sigma\nu}(Q-K)_\beta] \\
    N^{00}&=-8k_0^2-8kqcos\theta
\end{align}
where $\frac{1}{2}$ is symmetry factor. With hard thermal loop $T \sim p \gg q$, the self-energy in the first order of $\partial$ is
\begin{align}
    \Pi_{ar}^{00,AB(1)}&=C_Aig^2\delta^{AB}\int\frac{d^4K}{(2\pi)^3}\epsilon(k_0)\delta(K^2)\beta^3\sigma_{ij}k_ik_j\tilde{\chi}_g\frac{2k_0^2+2kqcos\theta}{p_0q_0-pqcos\theta+i\epsilon p_0+\frac{Q^2}{2}} \nonumber\\
    &=-C_Aig^2\delta^{AB}\beta^3 \frac{q_iq_j\sigma_{ij}}{2q^2}\int\frac{d^3k}{(2\pi)^3}(\frac{k^2(k+qcos\theta)(3cos^2\theta-1)}{kq_0-kqcos\theta+i\epsilon k+\frac{Q^2}{2}}-\frac{k^2(k+qcos\theta)(3cos^2\theta-1)}{kq_0+kqcos\theta+i\epsilon k-\frac{Q^2}{2}})\tilde{\chi}_g.
\end{align}
Then, we expand it by order $\frac{1}{k}$
\begin{align}
    \frac{k^2(k+qcos\theta)}{kq_0-kqcos\theta+i\epsilon k+\frac{Q^2}{2}}=\frac{k^2}{q_0-qcos\theta+i\epsilon}+(-\frac{Q^2}{2(q_0-qcos\theta+i\epsilon)^2}+\frac{qcos\theta}{q_0-qcos\theta+i\epsilon})k+O((\frac{1}{k})^0), \nonumber\\
    \frac{k^2(k+qcos\theta)}{kq_0+kqcos\theta+i\epsilon k-\frac{Q^2}{2}}=\frac{k^2}{q_0+qcos\theta+i\epsilon}+(\frac{Q^2}{2(q_0+qcos\theta+i\epsilon)^2}+\frac{qcos\theta}{q_0+qcos\theta+i\epsilon})k+O((\frac{1}{k})^0).
\end{align}
The first term in the order of $k^2 \sim T^2$ is
\begin{align}
    \Pi_{ar}^{00,AB(1)}=C_A ig^2 \delta^{AB}\beta^3 \frac{q_iq_j\sigma_{ij}}{2q^2}\int\frac{d^3k}{(2\pi)^3} (\frac{k^2(3cos^2\theta-1)}{q_0-qcos\theta+i\epsilon}-\frac{k^2(3cos^2\theta-1)}{q_0+qcos\theta+i\epsilon})\tilde{\chi}_g=0.
\end{align}
The particles and antiparticles cancel each other. The second term  in the order of $k \sim T$ is
\begin{align}
    \Pi_{ar}^{00,AB(1)}&=C_A ig^2 \delta^{AB}\beta^3 \frac{q_iq_j\sigma_{ij}}{2q^2}\int\frac{d^3k}{(2\pi)^3}(-\frac{Q^2(3cos^2\theta-1)}{2(q_0-qcos\theta+i\epsilon)^2}+\frac{qcos\theta(3cos^2\theta-1)}{q_0-qcos\theta+i\epsilon} \nonumber\\
    &-\frac{Q^2(3cos^2\theta-1)}{2(q_0+qcos\theta+i\epsilon)^2}-\frac{qcos\theta(3cos^2\theta-1)}{q_0+qcos\theta+i\epsilon})k\tilde{\chi}_g \nonumber\\
    &=-C_A \frac{\beta^3 q_iq_j\sigma_{ij}}{4\pi^2q^2} \delta^{AB} ig^2 \int dk k^3\tilde{\chi}_g[2(3x^2-1)-(3x^3-2x)ln(\frac{q_0+q+i\epsilon}{q_0-q+i\epsilon}).
\end{align}
\begin{figure}[htbp]
    \centering
    \includegraphics[width=0.3\textwidth]{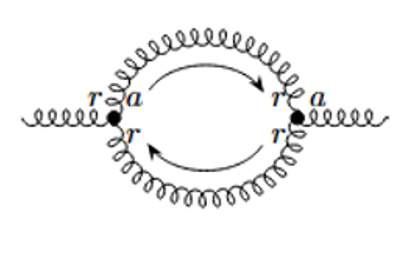}
    \label{fig:HTL3}
\end{figure}

The self-energy is
\begin{align}
    \Pi_{ar}^{\mu\nu,AB}&=\frac{1}{2}g^2f_{ADC}f_{BEF}\delta^{DE}\delta^{CF} \int \frac{d^4K}{(2\pi)^4}D_{ar}^{\alpha\beta}(K-Q)D_{rr}^{\sigma\rho}(K)[g_{\mu\alpha}(2Q-K)_\rho+g_{\alpha\rho}(2K-Q)_\mu-g_{\rho\mu}(K+Q)_\alpha] \nonumber\\
    &\times[g_{\nu\beta}K_\sigma+g_{\beta\sigma}(Q+K)_\nu+g_{\sigma\nu}(Q-2K)_\beta] \nonumber\\
    &=-\frac{1}{2}C_Aig^2 \int \frac{d^4K}{(2\pi)^4}\epsilon(k_0)\delta(K^2)(\frac{1}{2}+f_g)\frac{N^{\mu\nu}}{(q_0-k_0-i\epsilon)^2-(q-k)^2} \nonumber\\
    N_{\mu\nu}&=(P_T^{\alpha\beta}(K-Q)+\frac{(K-Q)^2}{(k-q)^2}u^\alpha u^\beta)[g_{\mu\alpha}(2Q-K)_\rho+g_{\alpha\rho}(2K-Q)_\mu-g_{\rho\mu}(K+Q)_\alpha] \nonumber\\
    &\times(P_T^{\sigma\rho}(K)+\frac{K^2}{k^2}u^\sigma u^\rho)[g_{\nu\beta}K_\sigma+g_{\beta\sigma}(Q+K)_\nu+g_{\sigma\nu}(Q-2K)_\beta] \\
    N^{00}&=-8k_0^2+8kqcos\theta
\end{align}
where $\frac{1}{2}$ is symmetry factor. With hard thermal loop $T \sim p \gg q$, the self-energy in the first order of $\partial$ is
\begin{align}
    \Pi_{ar}^{00,AB(1)}&=-C_Aig^2\delta^{AB}\int\frac{d^4K}{(2\pi)^3}\epsilon(k_0)\delta(K^2)\beta^3\sigma_{ij}k_ik_j\tilde{\chi}_g\frac{2k_0^2-2kqcos\theta}{p_0q_0-pqcos\theta+i\epsilon p_0-\frac{Q^2}{2}} \nonumber\\
    &=-C_Aig^2\delta^{AB}\beta^3 \frac{q_iq_j\sigma_{ij}}{2q^2}\int\frac{d^3k}{(2\pi)^3}(\frac{k^2(k-qcos\theta)(3cos^2\theta-1)}{kq_0-kqcos\theta+i\epsilon k-\frac{Q^2}{2}}-\frac{k^2(k-qcos\theta)(3cos^2\theta-1)}{kq_0+kqcos\theta+i\epsilon k+\frac{Q^2}{2}})\tilde{\chi}_g.
\end{align}
Then, we expand it by order $\frac{1}{k}$
\begin{align}
    \frac{k^2(k-qcos\theta)}{kq_0-kqcos\theta+i\epsilon k-\frac{Q^2}{2}}=\frac{k^2}{q_0-qcos\theta+i\epsilon}+(\frac{Q^2}{2(q_0-qcos\theta+i\epsilon)^2}-\frac{qcos\theta}{q_0-qcos\theta+i\epsilon})k+O((\frac{1}{k})^0), \nonumber\\
    \frac{k^2(k-qcos\theta)}{kq_0+kqcos\theta+i\epsilon k+\frac{Q^2}{2}}=\frac{k^2}{q_0+qcos\theta+i\epsilon}+(-\frac{Q^2}{2(q_0+qcos\theta+i\epsilon)^2}-\frac{qcos\theta}{q_0+qcos\theta+i\epsilon})k+O((\frac{1}{k})^0).
\end{align}
The first term in the order of $k^2 \sim T^2$ is
\begin{align}
    \Pi_{ar}^{00,AB(1)}=-C_A ig^2 \delta^{AB}\beta^3 \frac{q_iq_j\sigma_{ij}}{2q^2}\int\frac{d^3k}{(2\pi)^3} (\frac{k^2(3cos^2\theta-1)}{q_0-qcos\theta+i\epsilon}-\frac{k^2(3cos^2\theta-1)}{q_0+qcos\theta+i\epsilon})\tilde{\chi}_g=0.
\end{align}
The particles and antiparticles cancel each other. The second term  in the order of $k \sim T$ is
\begin{align}
    \Pi_{ar}^{00,AB(1)}&=-C_A ig^2 \delta^{AB}\beta^3 \frac{q_iq_j\sigma_{ij}}{2q^2}\int\frac{d^3k}{(2\pi)^3}(\frac{Q^2(3cos^2\theta-1)}{2(q_0-qcos\theta+i\epsilon)^2}-\frac{qcos\theta(3cos^2\theta-1)}{q_0-qcos\theta+i\epsilon} \nonumber\\
    &+\frac{Q^2(3cos^2\theta-1)}{2(q_0+qcos\theta+i\epsilon)^2}+\frac{qcos\theta(3cos^2\theta-1)}{q_0+qcos\theta+i\epsilon})k\tilde{\chi}_g \nonumber\\
    &=-C_A \frac{\beta^3 q_iq_j\sigma_{ij}}{4\pi^2q^2} \delta^{AB} ig^2 \int dk k^3\tilde{\chi}_g[2(3x^2-1)-(3x^3-2x)ln(\frac{q_0+q+i\epsilon}{q_0-q+i\epsilon}).
\end{align}
\begin{figure}[htbp]
    \centering
    \includegraphics[width=0.3\textwidth]{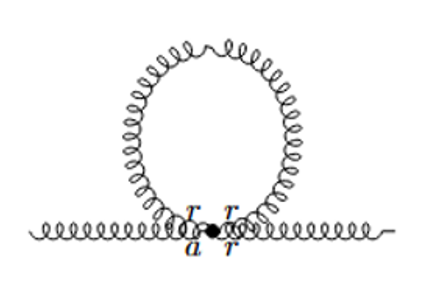}
    \label{fig:HTL4}
\end{figure}

The self-energy is
\begin{align}
    \Pi_{ar}^{\mu\nu,AB}&=\frac{ig^2}{2} \int \frac{d^4K}{(2\pi)^4}[f^{EAB}f^{ECD}(g^{\mu\alpha}g^{\nu\beta}-g^{\mu\beta}g^{\nu\alpha})+f^{EAC}f^{EDB}(g^{\mu\beta}g^{\nu\alpha}-g^{\mu\nu}g^{\alpha\beta}) \nonumber\\
    &+f^{EAD}f^{EBC}(g^{\mu\nu}g^{\alpha\beta}-g^{\mu\alpha}g^{\beta\nu})].
\end{align}
In the 00 component, the self-energy is
\begin{align}
    \Pi_{ar}^{00,AB(1)}&=C_A \delta^{AB}ig^2 \int \frac{d^4p}{(2\pi)^3} sgn(p_0) \beta^3 \frac{\sigma_{ij}p_i p_j}{2} \tilde{\chi}_g \delta(P^2) \nonumber\\
    &=C_A \delta^{AB}ig^2 \beta^3 \frac{q_i q_j \sigma_{ij}}{q^2} \int \frac{d^3p}{(2\pi)^3} \tilde{\chi}_g (3cos^2\theta-1)p \nonumber\\
    &=0.
\end{align}
Summing all diagrams, the self-energy in the first order of $\partial$ is 
\begin{align}
    \Pi_{ar}^{00,AB(1)}&=-i \delta^{AB}\beta \delta m^2 A^{00}, \nonumber\\
    A^{00}&=\frac{q_i q_j \sigma_{ij}}{2q^2}[2(3x^2-1)-(3x^3-2x)ln(\frac{q_0+q+i\epsilon}{q_0-q+i\epsilon})],  \nonumber\\
    \delta m^2&=N_f \frac{g^2\beta^2}{\pi^2} \int dp p^3 \tilde{\chi}_q+C_A \frac{g^2\beta^2}{\pi^2} \int dp p^3 \tilde{\chi}_g.
\end{align}
We use the solution of $\tilde{\chi}_{q/g}$
\begin{align}
        \tilde{\chi}_{q}&=-7.75sgn(p_0) \beta^{-\alpha_q} p^{-\alpha_q} \frac{\eta}{s} f_q^{(0)}(1-f_q^{(0)}) \nonumber\\
    \tilde{\chi}_{g}&=-4.5sgn(p_0) \beta^{-\alpha_g} p^{-\alpha_g} \frac{\eta}{s} f_g^{(0)}(1+f_g^{(0)}).
\end{align}
The value of $\delta m^2$ is
\begin{align}
    \int dp p^3 \tilde{\chi}_q&=-7.75 \frac{\eta}{s}\beta^{-0.62} \int dp \frac{e^{\beta p}p^{2.38}}{(e^{\beta p}+1)^2}=-19.38(\frac{1}{\beta})^4,  \nonumber\\
    \int dp p^3 \tilde{\chi}_g&=-4.5 \frac{\eta}{s}\beta^{-0.62} \int dp \frac{e^{\beta p}p^{2.38}}{(e^{\beta p}-1)^2}=-18.25(\frac{1}{\beta})^4,  \nonumber\\
    \delta m^2&=- \frac{g^2}{\pi^2 \beta^2}\frac{\eta}{s}(19.38N_f+18.25C_A).
\end{align}

\section{The Yukawa potential for shear in coordinate space }
The potential is 
\begin{align}
    V(r)&=C_F \frac{g^2}{(2\pi)^3}\int d^3q \frac{1}{q^2+\Pi_{L}(q_0=0)}e^{-i\vec{q}\cdot\vec{r}} \nonumber\\
    &=C_F \frac{g^2}{(2\pi)^3}\int d^3q \frac{1}{q^2+m_D^2+\beta \delta m^2 \frac{q_iq_j\sigma_{ij}}{q^2}}e^{-i\vec{q}\cdot\vec{r}} \nonumber\\
    &=C_F \frac{g^2}{(2\pi)^3}\int d^3q (\frac{1}{q^2+m_D^2}-\frac{\beta \delta m^2 q_iq_j\sigma_{ij}}{q^2(q^2+m_D^2)^2})e^{-i\vec{q}\cdot\vec{r}}.
\end{align}
Here, we expand the longitudinal self-energy in gradients, with $\Pi_L^{(0)}$ and $\Pi_L^{(1)}$ being $m_D^2$ and $\beta \delta m^2\frac{q_iq_j\sigma_{ij}}{q^2}$, respectively. We take that $r$ is parallel with $z$ axial $\vec{r}=(0,0,r)$ and the integration can be reduced as 
\begin{align}
    \int d^3 q \frac{ q_iq_j\sigma_{ij}}{q^2(q^2+m_D^2)^2}e^{-i\vec{q}\cdot\vec{r}}&=-\frac{\partial}{\partial r_i} \frac{\partial}{\partial r_j}\int d^3 q \frac{\sigma_{ij}}{q^2(q^2+m_D^2)^2}e^{-iqrcos\theta} \nonumber\\
    &=-2\pi^2 \sigma_{ij}\frac{\partial}{\partial r_i} \frac{\partial}{\partial r_j}\frac{2-e^{-m_D r}(2+m_D r)}{2m_D^4 r} \nonumber\\
    &=-2\pi^2\sigma_{ij}[\delta_{ij}\frac{m_D^2r^2e^{-m_Dr}+2m_Dre^{-m_Dr}+2e^{-m_Dr}-2}{2m^4_Dr^3} \nonumber\\
    &-r_ir_j(\frac{e^{-m_D r}}{2m_Dr^2}+\frac{3(e^{-m_D}-1)}{m_D^4 r^5}+\frac{3e^{-m_D r}}{m_D^3 r^4}+\frac{3e^{-m_D r^3}}{2m_D^2 r})].
\end{align} 
The potential for shear in coordinate space is
\begin{align}
    V(r)&=-C_F \frac{g^2}{4\pi}\frac{e^{-m_Dr} }{r}-C_F\frac{g^2\beta \delta m^2}{4\pi}\frac{r_ir_j\sigma_{ij}}{r^2}(\frac{e^{-m_D r}}{2m_D}+\frac{3(e^{-m_D}-1)}{m_D^4 r^3}+\frac{3e^{-m_D r}}{m_D^3 r^2}+\frac{3e^{-m_D r}}{2m_D^2 r}) .
\end{align}

\section{The shear correction to the perturbative potential in the KMS model}
In this section, we will demonstrate how the anisotropic Debye mass is used to derive the shear-corrected perturbative potential for the KMS model. The anisotropic Debye mass for shear is
\begin{align}
    m^2&=m_D^2-\beta \delta m^2 \frac{q_i q_j \sigma_{ij}}{q^2}, \\
    m&=m_D+ \triangle m=m_D \sqrt{1-\frac{\beta\delta m^2}{m_D^2}\frac{q_iq_j\sigma_{ij}}{q^2}} \approx m_D-\frac{\beta\delta m^2}{2m_D}\frac{q_iq_j\sigma_{ij}}{q^2},
\end{align}
where $\triangle m=-\frac{\beta\delta m^2}{2m_D}\frac{q_iq_j\sigma_{ij}}{q^2}$. The Karsch-Mehr-Satz model is
\begin{align}
    V_{KMS}(r)=-\frac{\alpha}{r}(1+m_D r)exp(-m_D r)+2 \frac{\sigma}{m_D}[1-exp(-m_D r)]-\sigma r exp(-m_D r).
\end{align}
Performing Fourier transform, the potential function in momentum space can be obtained
\begin{align}
    V_{KMS}(q)=-\frac{4\pi\alpha}{q^2+m^2}-\frac{8\pi(\alpha m^2+2\sigma)}{(q^2+m^2)^2}+\frac{8\pi\sigma(q^2-3m^2)}{(q^2+m^2)^3}+\frac{2\sigma}{m}(2\pi)^3\delta^3(q).
\end{align}
We perform a first-order expansion of the momentum-space potential $V(q)$ around $m_D$:
\begin{align}
    V_{KMS}(q)&=V_{KMS}(q)+\triangle V_{KMS}(q)=V_{KMS}(q)+\frac{\partial V}{\partial m} \mid_{m_D} \triangle m \\
    \triangle V_{KMS}(q)&=\frac{\beta \delta m^2}{m_D}\frac{q_i q_j \sigma_{ij}}{q^2}[\frac{4\pi\alpha m_D}{(q^2+m^2_D)^2}-\frac{16\pi m_D(\alpha m_D^2+2\sigma)}{(q^2+m^2_D)^3}+\frac{48\pi\sigma m_D(q^2-m_D^2)}{(q^2+m^2_D)^4}+\frac{\sigma}{m_D^2}(2\pi)^3\delta^3(q)].
\end{align}
By performing a Fourier transform, the first-order gradient potential function in coordinate space is obtained as
\begin{align}
    \triangle V_{KMS}(r)&=\frac{\sigma_{ij}r_i r_j}{r^2} f_{KMS}(r)     ,\nonumber\\
    f_{KMS}(r)&=2\pi \beta \delta m^2 [-\alpha \frac{A}{m^4_D r^3}+(\alpha m_D^2-\sigma)\frac{B}{m^6_D r^3}+\sigma \frac{C}{m_D^7 r^4}] ,\nonumber\\
    A&=6-e^{-m_D r}(m_D^3 r^3 +3m_D^2 r^2+6m_D r+6), \nonumber\\
    B&=24-e^{-m_D r}(m_D^4 r^4+4m_D^3 r^3+12m_D^2 r^2+24m_D r+24), \nonumber\\
    C&=144-e^{-m_D r}(m_D^5 r^5+6m_D^4 r^4+24m_D^3 r^3+72m_D^2 r^2+144m_D r+144).
\end{align}

\end{document}